\documentclass[pra,showpacs,floatfix,twocolumn]{revtex4-2}
\usepackage{graphicx}
\usepackage[utf8]{inputenc}
\usepackage{dcolumn}
\usepackage{bm}
\usepackage{amsmath}
\usepackage{amssymb}
\usepackage{wasysym}
\usepackage{color}
\usepackage{tabularx}
\usepackage{bm}

\begin{document}

\title{Guided Vortex Bullets}
\author{Carlos F. S\'{a}nchez$^{1,2}$, \'{A}ngel Paredes$^{2}$, Humberto
Michinel$^{2}$, Boris A. Malomed$^{3,4}$, and Jos\'{e} R. Salgueiro$^{2}$}
\affiliation{$^1$Grupo de Investigaci\'on en F\'isica (GIF), Universidad Polit\'ecnica
Salesiana, Cuenca, 010105, Ecuador.\\
$^2$Instituto de F\'isica e Ciencias Aeroespaciais (IFCAE), Universidade de
Vigo, 32004 Ourense, Spain.\\
$^3$Department of Physical Electronics, School of Electrical Engineering,\\
Faculty of Engineering, Tel Aviv University, Tel Aviv 69978, Israel. \\
$^{4}$Instituto de Alta Investigaci\'{o}n, Universidad de Tarapac\'{a},
Casilla 7D, Arica, Chile. }

\begin{abstract}

By means of the variational method and numerical simulations, we demonstrate
the existence of stable 3D nonlinear modes, \textit{viz}. vortex
\textquotedblleft bullets", in the form of pulsed beams carrying orbital
angular momentum, that can self-trap in a 2D waveguiding structure. Despite
the attractive self-interaction, which is necessary for producing the
bullets (bright solitons), and which readily leads to the collapse in the 3D
setting as well as to spontaneous splitting of vortex modes, we find a
critical value of the trapping depth securing the stabilization of the
vortex bullets. We identify experimental conditions for the creation of
these topological modes in the context of coherent optical and matter waves.
Collisions between the bullets moving in the unconfined direction are
found to be elastic. These findings contribute to the understanding of
self-trapping in nonlinear multidimensional systems and suggest new
possibilities for the stabilization and control of 3D topological solitons.

\end{abstract}

\maketitle

{} \textit{Introduction-.} Three-dimensional (3D) solitons~\cite{malomed2005spatiotemporal}, 
commonly referred to as \emph{wave/optical
bullets}~\cite{silberberg1990collapse} and modeled by generalized nonlinear
Schr\"{o}dinger equations (NLSEs)~\cite{fibichnonlinear}, offer a
fundamental testbed for studying the dynamics of many coherent physical
systems, from optical communications~\cite{kivshar2003optical} to
Bose-Einstein condensates~\cite%
{perez1998bose,strecker2002formation,khaykovich2002formation}, and from
plasma physics~\cite{shimizu1972automodulation} to cosmic structures~\cite%
{paredes2016interference}. These types of 3D solitary waves exist as
stationary states due to the balance between linear dispersive/diffractive
effects and nonlinear self-focusing, but they are usually prone to the
collapse (catastrophic self-compression). As well as the modulational
instability \cite{akhmediev1992modulation}, the collapse is driven by the
self-focusing nature of the nonlinearity~\cite%
{bespalov1966,sulem2007nonlinear}. In addition to the collapse-induced
instability of the fundamental (structureless) bullets, ones with embedded
vorticity (which is characterized by the corresponding topological charge)
are vulnerable to spontaneous splitting.

In quantum-matter settings, such as Bose-Einstein condensates (BECs) in
ultracold bosonic gases, the complex wavefunction $\Psi $, which obeys the
Gross-Pitaevskii equation (GPE), represents the order parameter in the
framework of the mean-field (MF) approximation \cite{pitaevskii1961vortex}.
The trap confining the ultracold gas is represented by the GPE term with
potential $V(\mathbf{r})$, while the effect of inter-atomic collisions is
reduced, by the MF approximation, to the cubic term in GPE. The coefficient
in front of the latter term is proportional to the respective scattering
length (positive or negative, for repulsive and attractive interactions
respectively) \cite{dalfovo1999theory}. In optics, the structure of
a linearly polarized electromagnetic pulsed beam, propagating in a nonlinear
graded-index waveguide, is confined in the transverse directions by a
linear refractive-index distribution that plays the role of the trapping
potential, while the optical Kerr effect gives rise to the cubic
self-focusing or defocusing depending on its sign~\cite{sulem2007nonlinear}. 
The two-dimensional transversal potential, if it is strong enough, contributes 
to the stabilization of the spatial soliton bullets in the case when widths of 
the soliton and potential  are similar. In fact, fundamental 
soliton bullets where already reported for parabolic~\cite{Yu1995} and 
periodic~\cite{Mihalache2004} potentials. Also, in a parabolic potential, 
similar solutions were described in Ref.~\cite{Raghavan2000} though for a 
system with the defocusing nonlinearity and normal dispersion.

As mentioned above, imparting the vorticity to nonlinear waves which is
quantified by the topological charge \cite{nye1974dislocations,coullet1989},
gives rise to much richer dynamics at the cost of fueling more instability
due to the growth of the splitting perturbations patterned along the azimuthal
coordinate~\cite{kruglov1985spiral}. Therefore, the realization of such
fascinating modes is a highly challenging~problem with many ramifications
\cite{hang2013ultraslow,Malomed2022}, and
precise control over the input wave and the balance of competing linear and
nonlinear effects are crucially important for achieving sustained stable
propagation~\cite{porras2019upper,chen1997self,zhang2022stabilization}.

In this work, we present the first, to the best of our knowledge,
prediction of the existence of stable 3D vortex bullets sustained and guided
by a combination of a shallow 2D trapping potential and self-attractive
cubic nonlinearity (2D lattice potentials can maintain stable 3D composite
solitary modes, built as a set of four density peaks with the
vorticity represented by phase shifts between them\cite{Leblond2007}). 
Our starting point is a (3+1)D NLSE with the temporal ($t$) and
spatial ($\mathbf{r}$) coordinates, which governs the propagation of a
coherent wave of amplitude $\Psi (\mathbf{r,t)}$ in the system with the cubic
self-focusing term. An axisymmetric trapping potential, $V(x,y)$, is defined
in the plane transverse to the propagation axis $z$, along which the wave
packet propagates freely. We adopt a Gaussian potential $V(x,y)$, which is a
commonly used experimental tool in BEC and optics.

By means of a variational approximation (VA) we predict approximate
eigenstates of the corresponding NLSE. The VA solutions are then used as an
input to numerically solve the reduced axisymmetric problem by means of a
finite-differences scheme~\cite{salgueiro2007computation} and a globally
convergent Newton method. Finally, we apply the Vakhitov-Kolokolov (VK)
necessary-stability criterion~\cite{Vakhitov1973}, and verify the
full stability by simulating the perturbed evolution with the help of the standard 
beam propagation method.

\textit{The model-.} We start by considering the GPE with the attractive
cubic term and a shallow Gaussian trapping potential with the axial
symmetry, $V(\rho )=-\gamma \exp \left( -\rho ^{2}/2\right) $, where $\left(
\rho ,\varphi \right) $ are the polar coordinates in the $(x,y)$ plane. The 
respective GPE takes the usual scaled form,

\begin{equation}
i\frac{\partial \Psi }{\partial t}+\nabla ^{2}\Psi -V(\rho )\Psi +|\Psi
|^{2}\Psi =0,  \label{eq:NLSE}
\end{equation}%
where $\nabla ^{2}$ is the kinetic-energy operator acting on coordinates $%
\left( x,y,z\right) $. The only free parameter in Eq. (\ref{eq:NLSE}), which
cannot be fixed by rescaling, is the strength $\gamma >0$ of the trapping potential
$V(\rho )$. The same equation governs the propagation of a linearly
polarized laser pulse in a cylindrical gradient-index waveguide with the
focusing Kerr nonlinearity and anomalous group-velocity dispersion. In the
latter case, $\gamma $ is proportional to the maximum value of the local increase 
of the refractive index, $t$ is the
propagation distance, and $z$ is the temporal coordinate in the reference
frame moving with the group velocity of the carrier wave. Equation (\ref%
{eq:NLSE}) conserves the norm $N=\int |\Psi |^{2}d^{3}\mathbf{r}$, angular
momentum $M_{z}=-i\int \Psi ^{\ast }\partial _{\varphi }\Psi d^{3}\mathbf{r}$,
 and the Hamiltonian:
\begin{equation}
H=\int \left[ |\vec{\nabla}\Psi |^{2}-\frac{1}{2}|\Psi |^{4}-\gamma e^{-\rho
^{2}/2}|\Psi |^{2}\right] d^{3}\mathbf{r}\,.  \label{eq:H}
\end{equation}%

We consider stationary localized states carrying integer vorticity
(topological charge) $\ell $ and real chemical potential $-\beta $ (in
optics, $\beta $ is the propagation constant):
\begin{equation}
\Psi (x,y,z,t)=\psi (\rho ,z)e^{i(\ell \varphi +\beta t)},
\label{eq:stationary_state}
\end{equation}%
for which the virial identity~\cite{Mihalache2004} can be derived from Eq. (\ref{eq:NLSE}):

\begin{equation}
H=-\beta \,N+\frac{1}{2}\int |\psi |^{4}d^{3}\mathbf{r}\,.  \label{virial}
\end{equation}%

\textit{The variational approximation (VA)-.} In this work we fix on
fundamental vortex modes, with $\ell =1$, which may be approximated by the
real variational ansatz for $\psi $, with transverse and axial
(longitudinal) widths parameters, $R$ and $\eta $:

\begin{equation}
\psi _{\mathrm{VA}}=\sqrt{\frac{N}{2\pi \eta }}\frac{1}{R^{2}}\,\rho \,\exp
\left( -\frac{\rho ^{2}}{2R^{2}}\right) \,\mathrm{sech}\left( \frac{z}{\eta }%
\right) \,.
\end{equation}%
Minimizing $H$ with respect to $R$ and $\eta $ for fixed $N$, a
straightforward calculation produces the following VA results:

\begin{equation}
\eta =\frac{16\pi R^{2}}{N}\,,\qquad N=\frac{16\pi \sqrt3 R}{(R^2+2)^\frac32} \sqrt{(R^2+2)^3-4\gamma R^4}\,,  \label{tauR}
\end{equation}%
the respective Hamiltonian per particle being

\begin{equation}
\frac{H}{N}=\frac{1}{R^{2}}-\frac{8\gamma }{(R^{2}+2)^{3}}\,,  \label{H/N}
\end{equation}%
and $\beta$ is calculated from Eq.~(\ref{virial}) by inserting the obtained variational 
wavefunction. Thus, for a given trapping-potential strength $\gamma $, the VA predicts the
existence of a family of stationary states parameterized by the
variational radius $R$, while the respective chemical potential can be
obtained from Eq. (\ref{virial}). Analyzing the respective Hessian matrix of the 
derivatives, we find that the VA solutions correspond to a local minimum of
the Hamiltonian (hence, they may be stable) if and only if $dN/dR<0$. Consequently, the 
domain of  \textit{potentially stable} states can be found by requiring that $N(R)$ [Eq.~(\ref{tauR})] has an inflection 
point satisfying $N'(R)=N''(R)=0$, which leads to $\gamma= 8/3$, $R=\sqrt{2}$. On the other hand there is also a 
limit value of $\gamma$ for which there exist solutions with $N\to 0$ and nonzero values of $R$ 
($\eta$ diverging as $N\to 0$). This limit value of $\gamma$ can be found 
by requiring that the expression under the square root in Eq.~(\ref{tauR}) and its derivative vanish 
simultaneously, which leads to $\gamma = 27/8$ and $R=2$.

\begin{figure}[th]
\begin{center}
\includegraphics[width=\columnwidth]{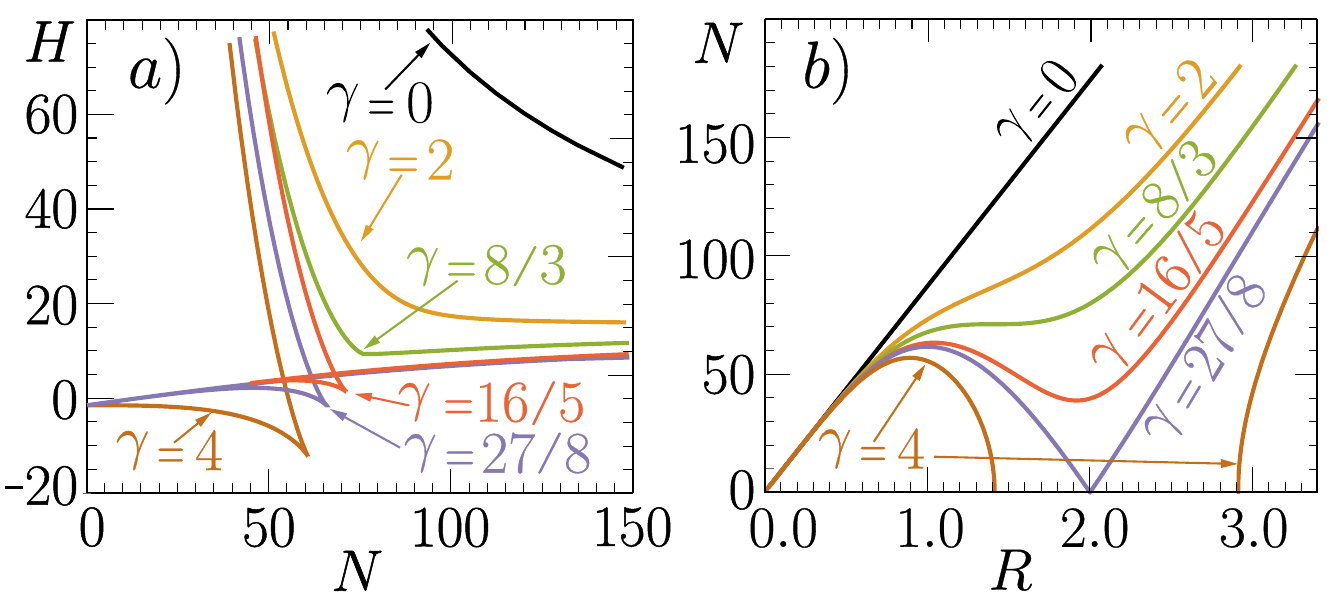}
\end{center}
\caption{a) Hamiltonian $H$ vs. norm $N$ \ for several values of the
potential strength $\protect\gamma $, as predicted by the VA. b) The
VA-predicted dependences $N$ on the transverse bullet's radius $R$, for the
same set of values of $\protect\gamma $.}
\label{fig:fig1}
\end{figure}

In Fig.~\ref{fig:fig1}(a) we plot the VA-predicted dependences $H(N)$ for different values of $\gamma$. 
As can be seen in the picture, the cust catastrophe is present for $\gamma>8/3$.
The respective dependences $N(R)$ are plotted Fig. \ref{fig:fig1} (b). 

\textit{Numerical solutions-.} Once the VA solution has been found it is
necessary to compare it with a numerical counterpart. Stationary states are
solutions of the following eigenvalue problem for the real amplitude,
produced by substitution $\psi (\rho ,z)=\rho ^{-1/2}u(\rho ,z)$:

\begin{equation}
\frac{\partial ^{2}u}{\partial z^{2}}+\frac{\partial ^{2}u}{\partial \rho
^{2}}-\frac{(\ell ^{2}-1/4)}{\rho ^{2}}u+\left[\frac{u^{2}}{\rho}-V(\rho )-\beta\right] u=0.  \label{eq:final}
\end{equation}%
Equation~(\ref{eq:final}) was solved by means of a finite-differences scheme~%
\cite{,salgueiro2007computation}, with the resulting nonlinear algebraic problem solved
iteratively using a globally convergent Newton method. This technique is
suitable, as the solutions have to meet the boundary condition $\psi (0)=0$
and thus $u(0)=0$, yielding a regular behavior at $\rho\rightarrow 0$, in spite
of the diverging factor $\rho ^{-1/2}$ in the above-mentioned substitution.

\begin{figure}[th]
\begin{center}
\includegraphics[width=\columnwidth]{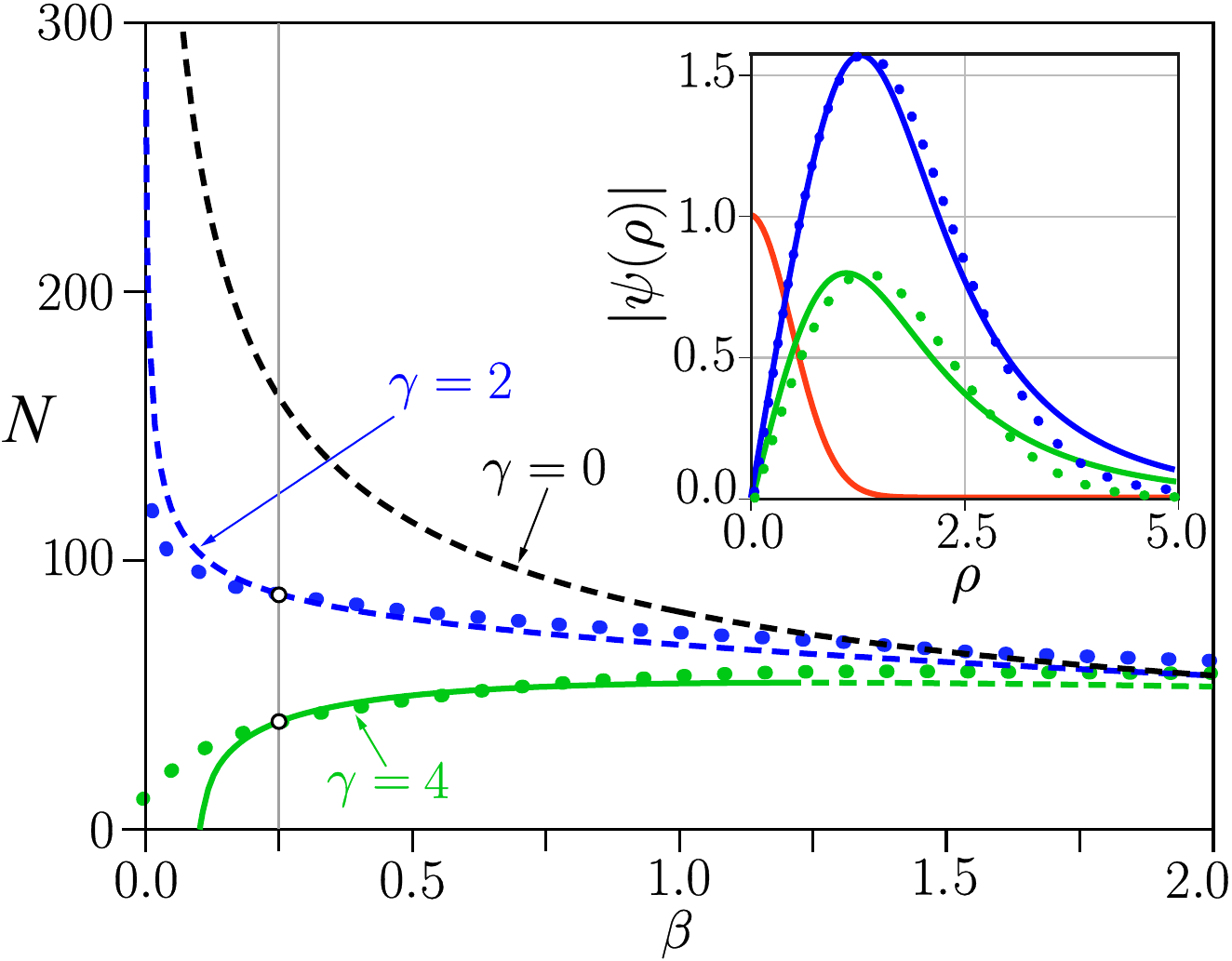}
\end{center}
\caption{Norm $N$ of the numerically found stationary vortex solutions vs.
the propagation constant $\protect\beta $ for different values of the
waveguide strength parameter $\protect\gamma $. Dashed and continuous lines 
indicate VK-unstable and VK-stable families respectively. The black (dashed) curve 
displays, for comparison, the family of unstable free-space solutions with 
$\protect\gamma =0$. Doted lines represent the solutions obtained with the 
variational method.   The inset shows numerical (solid) and variational (dotted) 
radial shapes of the solutions for 
$\protect\beta =0.25$ [designated by the vertical gray line in the main plot and 
with the colors corresponding to those of the $N(\protect\beta )$ curves], along 
with the Gaussian trapping profile $-V/\protect\gamma $ (red curve).  }
\label{fig:fig2}
\end{figure}

Dependences $N(\beta )$, found numerically for $\gamma =0$ (free space), $\gamma=2$ 
and $\gamma=4$ are displayed, along with their VA-predicted counterparts in Fig.~%
\ref{fig:fig2}. Naturally, the deeper the trap is, the fewer particles are
necessary to induce nonlinear self-trapping. Therefore, the values of $N$
are lower for stronger potentials with higher values of $\gamma $. The
free-space solitons, obtained for $\gamma =0$, are definitely unstable,
being included here for comparison. The continuous green ($\gamma =4$) and
dashed blue ($\gamma =2$) curves represent, respectively, the solution families
which are expected to be stable and unstable, according to the
Vakhitov-Kolokolov (VK) stability criterion, $dN/d\beta >0$, \cite{Vakhitov1973,malomed2005spatiotemporal,fibichnonlinear,sulem2007nonlinear}. 
The inset in Fig.~\ref{fig:fig2} displays the
Gaussian potential (the red curve) and stationary-solution profiles for $%
\beta =0.25$, designated by the vertical gray line in the main plot.

The blue $N(\beta )$ curve in Fig.~\ref{fig:fig2}, which corresponds to a
very shallow potential with $\gamma =2$, is qualitatively similar to
the free-space one for $\gamma =0$. The increase of the trapping depth to $%
\gamma >4$ (the green curve) dramatically changes the situation, generating
a cutoff value of the propagation constant, $\beta _{\min }\approx 0.104$,
that correspond to the linear eigenstate ($N=0$) confined by the trapping
potential in the $\left(x,y\right)$ plane. Another notable consequence of making 
the trap stronger is that the green $N(\beta )$ curve attains a (very flat) maximum around $\beta =1.24$, beyond
which the family is definitely VK-unstable, exhibiting $dN/d\beta <0$.

\begin{figure}[tbph]
\begin{center}
\includegraphics[width=\columnwidth]{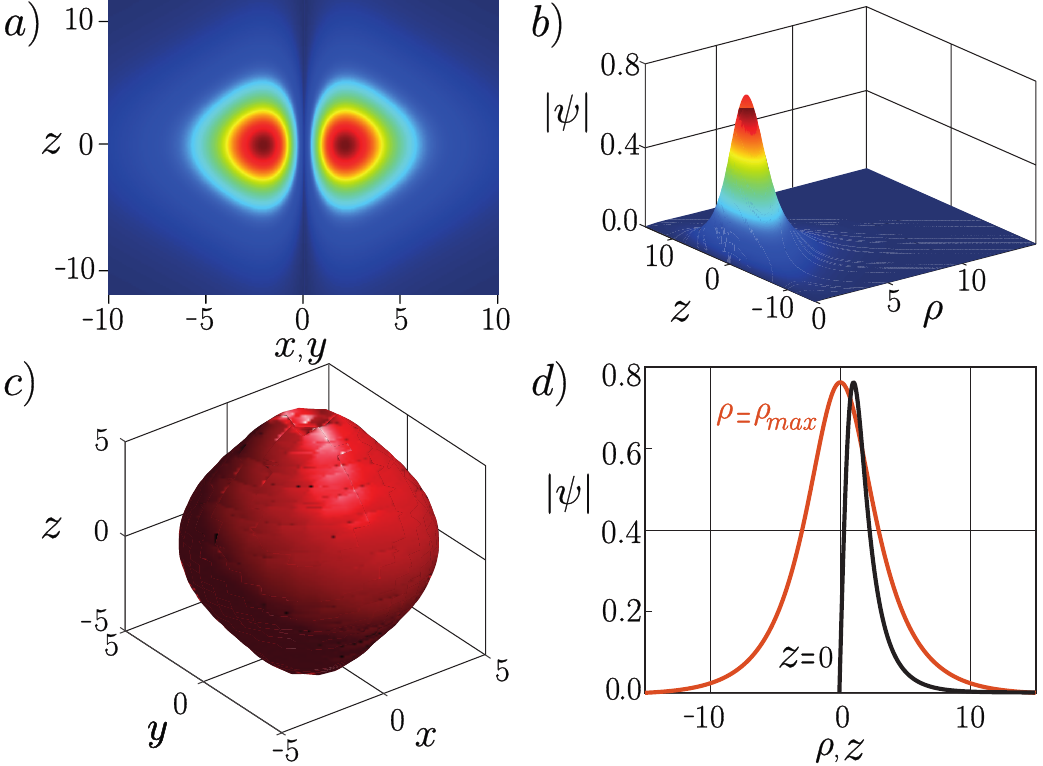}
\end{center}
\caption{A characteristic \emph{stable} numerical solution of Eq.~(\protect
\ref{eq:final}), with $\protect\gamma =4$, $\protect\beta =0.25$ and $\ell =1$. 
Displayed are contour and mesh views of $|\protect\psi |$, in a) and b),
respectively. Panel c) shows the 3D isosurface, and d) plots the radial and
axial (longitudinal) profiles of $|\protect\psi |$ at $z=0$ and $\protect%
\rho =\protect\rho _{\max }$ (at $\protect\rho $ realizing the maximum of $|%
\protect\psi |$).}
\label{fig:fig3}
\end{figure}

A characteristic example of the stationary vortex state is presented in Fig.~%
\ref{fig:fig3}, for $\gamma =4$ and $\beta =0.25$, displaying the contour
and mesh views  in a) and b) respectively, of $\left\vert \psi \left(
x,y,z\right) \right\vert $ in plane $\left( \rho ,z\right) $, 
along with the 3D isosurface view in c) and the cross-section
profiles in d) at $z=0$ and $\rho =\rho _{\max }$ (the value of $\rho $ at
which $|\psi |$ attains its maximum). This solution is a
stable one as shown below.

\textit{The stability-.} Because the VK criterion is only necessary but not
sufficient for the full stability (particularly it cannot detect the
plausible splitting instability of vortex solitons\cite{malomed2005spatiotemporal,Malomed2022}), 
final conclusions about the stability of stationary states
satisfying the VK criterion were made by means of direct simulations of
their perturbed propagation.  The initial condition determines the topology of the emerging state 
provided that it is a stable one. This fact is important because it suggests that nontrivial 
vortex states may be experimentally created by means of choosing the appropriate input.

\begin{figure}[tbh]
\begin{center}
\includegraphics[width=\columnwidth]{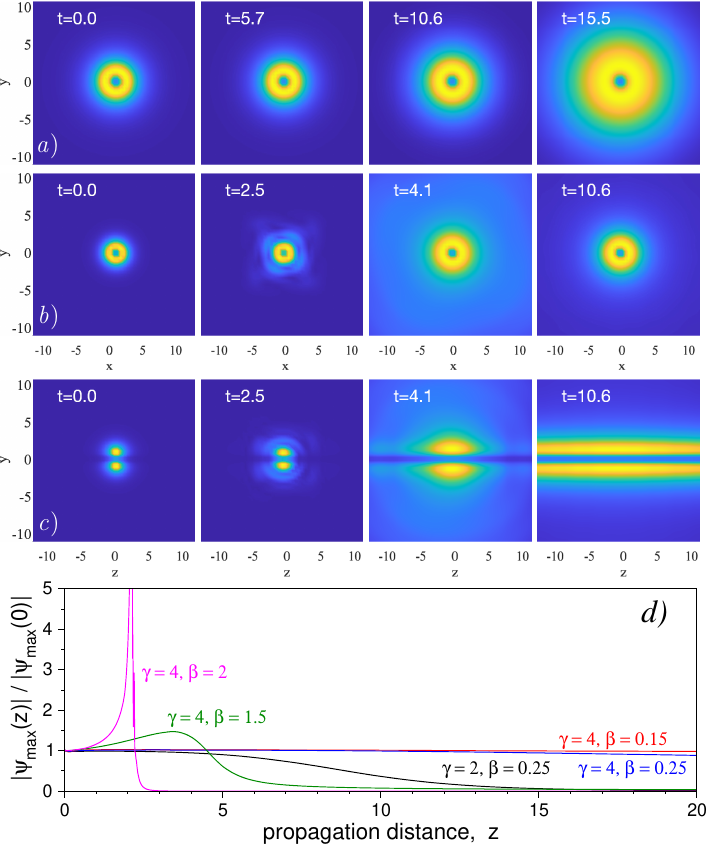}
\end{center}
\caption{The simulated perturbed evolution of two unstable solutions for a) $%
\protect\gamma =2$, $\protect\beta =0.25$ and b) $\protect\gamma =4$, $%
\protect\beta =2$, plotted in the $(x,y)$ plane. c) The same as in b), but
in the $(z,y)$ plane. d) The evolution of the field's amplitude (normalized
to its value at $t=0$) for different solutions, as indicated (both
stable, for $\protect\gamma =4$, $\protect\beta =0.15$ and  $\beta=0.25$ and
unstable for all other cases).}
\label{fig:simulations}
\end{figure}

First, it was corroborated that all solution
families which do not satisfy the VK criterion are indeed unstable, such as
those in Fig.~\ref{fig:fig2} corresponding to $\gamma =0$ and $\gamma=2$, as well
as to $\gamma =4$ and $\beta >1.24$. Characteristic examples of the unstable
evolution are displayed in Fig. \ref{fig:simulations}. Panel (a) represents a simulation 
belonging to $\gamma =2$ showing the 
decaying (spreading out) evolution of the state.  On the other hand, the evolution of 
a solution with $\gamma =4$, in the VK-unstable region of the negative slope 
($\beta =2$) is shown in panels (b) and (c). In this case the bullet, after an initial contraction 
which results in a collapse as is $\beta$ large enough, follows a spreading out behavior 
(actually, the post-collapse evolution may be irrelevant, as the usual GPE may 
become invalid for a strongly collapsed state). At this point the bullet is stabilized in 
the $(x,y)$ plane [see panel (b)] due to the action of the trapping potential, but is 
unstable in the axial direction, as shown in panel (c).


The evolution of the maximum amplitude for different solutions is shown in
Fig. \ref{fig:simulations}(d). In accordance to what is said above, the unstable
solutions for $\gamma =4$, \textit{viz}. are destroyed by the initial contraction 
($\beta=1.5$) or collapse ($\beta=2$). On the other hand, the instability of the
mode with $\gamma =2$ and $\beta =0.25$ eventually does not lead to the
collapse, but to decay.

The systematic simulations demonstrate that the vortex solitons which belong
to the VK-stable family, such as the one with $\gamma =4$ and $\beta <1.24$,
represented by the continuous green branch in Fig. \ref{fig:fig2}, are
indeed fully stable. Examples, plotted in Fig. \ref{fig:simulations}(d) for $\gamma
=4$ and $\beta =0.15$ and $\beta=0.25$, corroborate the stability -- at least, up
to $t=200$ which, roughly speaking, corresponds to $20$ characteristic
diffraction times for these modes. The conclusion is that the potential
should be sufficiently strong (approximately, with $\gamma \gtrsim 3.1$) for 
maintaining the stability of the trapped vortex modes, and their propagation 
constant should not be too large.

\begin{figure}[tbh]
\begin{center}
\includegraphics[width=\columnwidth]{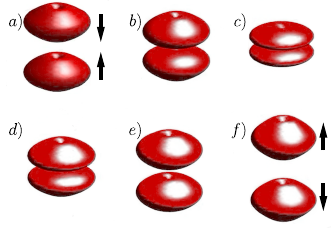}
\end{center}
\caption{The elastic head-on collision of two stable vortex bullets from\
Fig.~\protect\ref{fig:fig2} with opposite signs, which are set in motion
along the $z$ axis with opposite initial velocities. 
The collision
leads to the elastic rebound of the solitons. The frames a) to f) correspond,
respectively, to evolution times $t=0.0$, $7.6$, $15.2$, $22.9,30.5,$ and $%
38.1$.}
\label{fig:collision}
\end{figure}

Once stable vortex bullets are found, it is natural to test their robustness
against mutual collisions in the unconfined direction ($z$), cf. Ref.\cite{Leblond2009}. 
Simulations demonstrate in Fig.~\ref{fig:collision} that the
repulsive head-on collision of two stable vortex bullets from Fig.~\ref%
{fig:fig3}, with opposite initial velocities, identical vorticities, $\ell =1
$, and the phase shift $\pi $ (opposite signs) seems completely elastic,
which is a typical outcome. In-phase collisions between the vortex bullets
with identical signs are elastic too, leading to their mutual passage
provided that the initial velocities are large enough (not shown here). 

\textit{Conclusion.-} In this work we have studied the formation, stability,
and propagation dynamics of bullets (3D solitons), carrying vorticity $\ell=1$,
in the  3D cubic self-focusing medium with the 2D axisymmetric trapping
Gaussian potential. By means of the variational method and systematic
numerical analysis, we have demonstrated, for the first time to our
knowledge, the existence of stable vortex-carrying wavefunctions under
experimentally accessible conditions, that may be realized in BEC and
optics. The stability area for the vortex bullets, and the dependence of
their shape on the control parameters (the depth of the trapping potential
and the bullet's norm) have been identified. Elastic collisions between the
stable bullets moving in the unconfined direction have been demonstrated.
Unstable bullets demonstrate either intrinsic collapse or spontaneous decay.

As an extension of the work it may be relevant to consider the bullets with
multiple embedded vorticity, and also study collisions between ones with
opposite vorticities. Also interesting is to extend the analysis for media
with competing nonlinearities, which may be relevant for optics\cite{quiroga1997} 
and BEC\cite{Dong2024} alike. Finally, a study of higher-charge vortices is 
interesting because, although it is known that they exhibit stronger 
azimuthal-modulation (splitting) 
instabilities, there is no universal topological argument that forces vortex 
bullets with charge $\ell > 1$ to be unstable. Several models with competing 
nonlinearities are known to support stable higher-charge 
vortices\cite{michinel2004square}. Exploring whether suitable trapping 
configurations may stabilize $\ell > 1$ bullets in this setting is an interesting 
direction for further research.

Datasets generated for the figures are accesible (see Ref.~\cite{Michinel2025}).

\section*{Acknowledgements}

This publication is part of the R\&D\&i project PID2023-146884NB-I00 funded 
by MCIN/AEI/10.13039/501100011033/. This work was also supported by 
grants ED431B 2021/22 and GPC-ED431B 2024/42 (Xunta de Galicia).


%

\end{document}